\documentclass[preprint2]{aastex}
\newcommand{\comments}[1]{}

\shorttitle{Astrometric companions to nearby stars}
\shortauthors{Tokovinin, Hartung, Hayward, Makarov}

\begin{document}

\title{Revealing   companions  to   nearby   stars  with   astrometric
  acceleration \footnote{Based on  observations obtained at  the Gemini
    Observatory (Program ID GS-20011B-Q-69). }
%
}

\author{Andrei Tokovinin}
\affil{Cerro Tololo Inter-American Observatory, Casilla 603, La Serena, Chile}
\email{atokovinin@ctio.noao.edu}

\author{Markus Hartung, and Thomas L. Hayward }
\affil{Gemini Observatory, Southern Operations Center,  Casilla 603, La Serena, Chile}
\email{mhartung@gemini.edu, thayward@gemini.edu}

\author{Valeri V. Makarov}
\affil{US Naval Observatory, 3450 Massachusetts Sve. NW, Washington, DC, 20392-5420}
\email{valeri.makarov@usno.navy.mil}


\begin{abstract}
A subset of  51 Hipparcos astrometric binaries among  FG dwarfs within
67\,pc  has been  surveyed with  the  NICI adaptive  optics system  at
Gemini-S,  directly  resolving for  the  first  time 17  sub-arcsecond
companions and 7 wider ones.  Using these data together with published
speckle  interferometry of  57  stars, we  compare  the statistics  of
resolved  astrometric  companions  with  those of  a  simulated  binary
population.   The fraction  of resolved  companions is  slightly lower
than expected  from  binary statistics.  About 10\% of astrometric
companions could  be ``dark''  (white dwarfs and  close pairs  of late
M-dwarfs).   To   our  surprise,  several  binaries   are  found  with
companions  too  wide to  explain  the  acceleration.  Re-analysis  of
selected  intermediate astrometric data  shows that  some acceleration
solutions in the original Hipparcos catalog are spurious.
\end{abstract}


\section{Introduction}
\label{sec:intro}

The Hipparcos
catalog  \citep{HIP1997d}  contains  objects  with  non-linear  proper
motions (PM) caused by binary companions; they are referred to as {\em
  acceleration} or $\dot{\mu}$ binaries. Their positions are described
by 2nd  (acceleration) or higher-order polynomials or,  in rare cases,
by  full orbital  solutions. In  addition, accelerated  motion  due to
companions in the  so-called $\Delta \mu$ binaries can  be revealed by
the difference  between the  Hipparcos PM measured  on a time  base of
3.2\,y  and the  long-term PM  from the  Tycho-2  \citep{TYC2} catalog
\citep[][hereafter  MK05]{MK05}, exploiting  a time  base of  almost a
century.   So far  little is  known  about both  types of  astrometric
binary systems, yet  they cover an important range  of orbital periods
from  a  few  to  a  few hundred  years  where  alternative  detection
techniques are not very efficient, especially for low-mass companions.
The goal of our study is  to get a better understanding of astrometric
companions  and  their parameters  and  to  use  this information  for
improving binary statistics.

A thorough knowledge of binary and multiple star statistics is needed for
the study of star formation, for stellar population synthesis, for
predicting the frequency of supernovae, blue stragglers, X-ray
binaries, etc. The statistical properties of binaries strongly depend on
stellar mass. Only for nearby solar-mass dwarfs, however, current
techniques (including Hipparcos astrometry) cover the discovery space
well enough to enable statistical completeness.  The classical work on
G-dwarf binaries by \citet{DM91} has been recently superseded by
\citet{Raghavan10} who revised the frequency of triple and
higher-order hierarchies from 5\% to 10\% even in this well-studied
sample within 25\,pc.  Given that there are only 56 hierarchical
stellar systems in this small volume of space, we need a much larger
sample for an un-biased statistical study of a multiplicity of 3 or
higher.

The  Hipparcos  catalog  is  complete  for dwarfs  more  massive  than
0.8\,$M_\odot$  with  parallax  larger  than 15\,mas  (the  number  of
objects within distance $d$ is  proportional to $d^3$). Hence, the sample of
$\sim$5000 FG  dwarfs within 67\,pc derived  from Hipparcos (hereafter
FG-67pc)  is   ideally  suited   for  statistical  study   of  stellar
hierarchies.  The radial  velocity (RV) has been measured  for a large
fraction of these  stars by the Geneva-Copenhagen Survey  of the Solar
neighbourhood,  GCS  \citep{N04},  revealing  short-period  binaries.
Wide companions can be retrieved by data mining \citep[e.g.][]{Tok11}.
Unfortunately,   the   parameters   of   astrometric   binaries   with
intermediate  periods from  few to  few hundred  years  remain largely
unknown.   The FG-67pc  sample  contains $N_\mu  =  329$ $\Delta  \mu$
binaries  and $N_a  =  244$ acceleration  binaries  from MK05.   These
groups  overlap and leave  a total  of $N_{\mu,a}=343$  objects.  Only
half of those were also  detected as spectroscopic binaries by GCS and
other authors.

Many  nearby dwarfs  are searched  for exo-planets.   Dark astrometric
companions do not degrade the  RV precision of ongoing surveys, unlike
binaries with a mass ratio $\sim  1$ where light of both companions is
mixed in  the spectrum in  uncontrolled proportion due to  guiding and
seeing.  It is known that close binaries host fewer planets than single
stars, but  rare exceptions to  this rule give valuable  insights into
planet formation.  Such is the  case of HIP~101966 with a 72-y $\Delta
\mu$  companion  and a  3.6-y  planet  \citep{Chauvin11}.  Some  other
exo-planet  candidates in astrometric  binaries turn  out to  be brown
dwarfs on low-inclination orbits,  e.g.  the outer planet in HIP~27253
\citep{Benedict10} or HIP~4311 \citep{Sahlmann11}.

In  order to  improve our  understanding of  astrometric  binaries, we
conducted a  ``snapshot'' survey  using Adaptive Optics  (AO) imaging.
High-resolution imaging can achieve the following goals:
\begin{itemize}
\item
  Characterize  targets for exo-planet  search. 
\item
 Confirm  or  refute   Hipparcos  detections  for  nearby  astrometric
 binaries, estimate their reliability.
 
\item
 Determine  companion masses  from relative  photometry  and estimate
  orbital  periods from projected  separations.  These  constraints on
  companions are much tighter  than those obtained from the astrometry
  alone.

\item
 Provide  first-epoch measurement of  companion positions  for future
  orbit calculation.
\end{itemize}

The    results    of    these    observations   are    presented    in
Section~\ref{sec:AO}.   High-resolution imaging of  additional objects
is retrieved  from recent speckle-interferometry  observations and the
combined sample  of 99 stars is  studied in Section~\ref{sec:speckle}.
In  Section~\ref{sec:sim} we  compare our  findings  with simulations,
trying  to  put some  constraints  on  the  statistics of  astrometric
companions.  The reanalysis of Hipparcos Intermediate Astrometric Data
(HIAD)  in   Section~\ref{sec:bogus}  shows  that   some  acceleration
solutions  of  Hipparcos are  spurious.   We  discuss  the results  in
Section~\ref{sec:disc}.

\section{AO observations and results}
\label{sec:AO}

\begin{figure*}[ht]
\epsscale{2.0}
\plotone{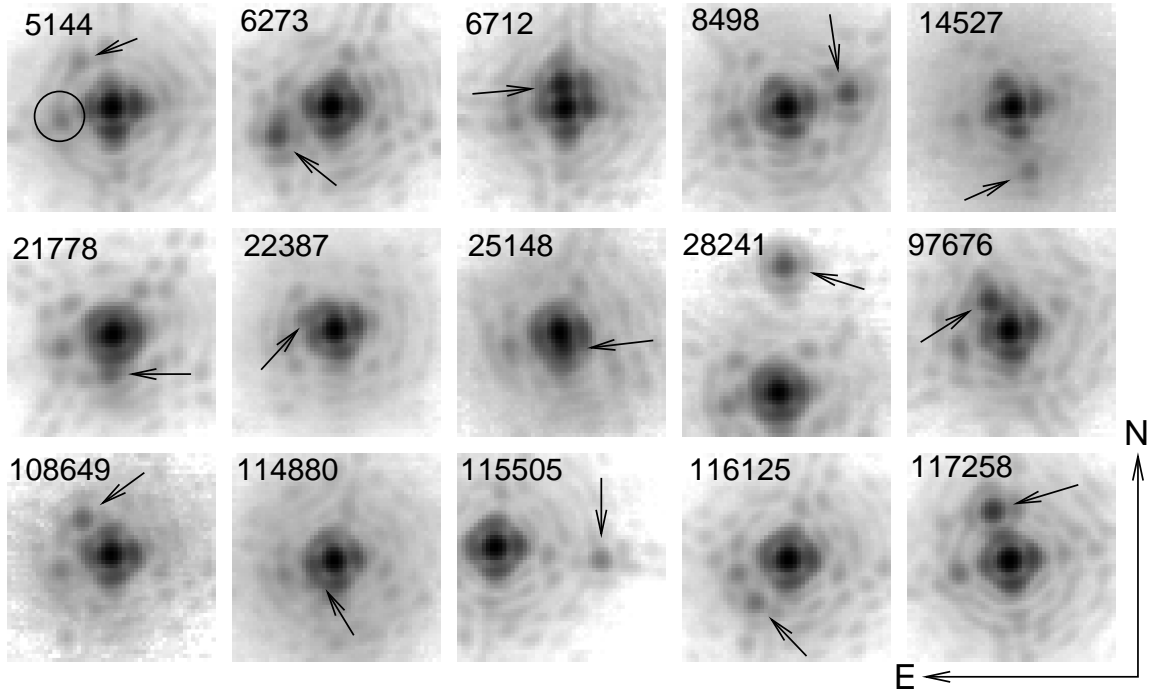}
\caption{Images of  some resolved close companions in  the red channel
  (2.272\,$\mu$m),   marked  by  with   the  HIP   numbers.   Negative
  logarithmic intensity scale from $10^{-3}$ to maximum, each fragment
  is 50x50 pixels ($0.9''$).  The prominent ``ghost'' companion to the
  left of  each target (circled in  the first image) is a reflex in the
  NICI optics.
\label{fig:images}
}
\end{figure*}

The Near-Infrared Coronagraphic Imager, NICI, on the Gemini South
telescope is an 85-element curvature adaptive optics (AO) instrument
based on natural guide stars \citep{NICI,Chun08}.  We used NICI in the
non-coronagraphic mode, as a classical AO system with simultaneous
imaging at two wavelengths.  The two detectors have $1024^2$ pixels of
18\,mas (milliarcseconds) size, covering a square field of $18''$.  To
avoid saturation,  we selected narrow-band filters with central
  wavelengths of 2.272\,$\mu$m and 1.587\,$\mu$m for the red and blue
  imaging channels.

The observations  of 51 Hipparcos  astrometric binaries were  taken in
queue mode in  the period from September 15 to  November 8, 2011 using
11.6\,h of  the 14.7\,h allocated  time.  The observing  procedure and
data reduction  are the  same as in  \citep[][hereafter THH10]{THH10}.
The images of  each target at 5 dither  positions were median-combined
after removing  bad pixels, subtracting  the median to remove  the sky
background, dividing by a flat field, and suitable shifts.

\begin{figure}[ht]
\epsscale{1.0}
\plotone{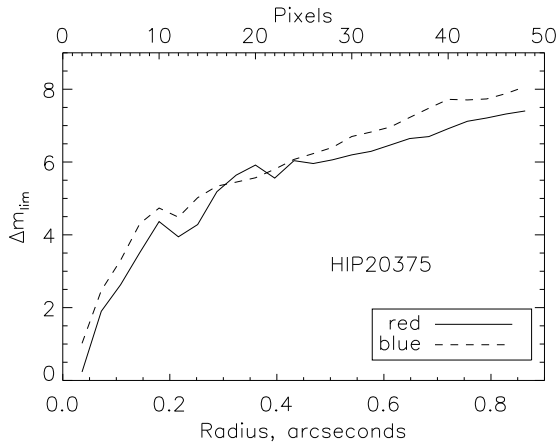}
\caption{Limiting magnitude difference for companion detection in the red (full
  line) and blue (dashed line) channels for HIP~20375, a typical case
  with Strehl ratio of 0.40 in the red channel.
 \label{fig:det}}
\end{figure}

A complete list of the observed stars is given in the next Section.  A
total  of 24 companions  with separations  from 0\farcs1  to 13\farcs8
were  resolved,  17  of   those  are  sub-arcsecond.   While  brighter
companions are  quite obvious, the  faint ones (e.g.   HIP~114880) are
buried    in   the    static    speckle   and    difficult   to    see
(Fig.~\ref{fig:images}).   The  speckle structure  is  dominated by  a
cross-like pattern in the first  diffraction ring and by several knots
along the diffraction  rays (the pupil mask of  NICI covers the spider
with oversized  stripes creating this  particular diffraction pattern).
The reality of detections is  checked by ``blinking'' the red and blue
images and by  comparing with other stars. Some  companions are better
seen in the blue images  where the speckle structure is less prominent
and point  sources are  sharper.  The faint  ``ghost'' with  $\Delta m
\sim 4.3$  at 0\farcs24 to  the left of  each star is produced  by the
NICI optics.

The limiting magnitude for companion detection was determined from the
intensity    fluctuations   in   annular    zones,   as    in   THH10.
Figure~\ref{fig:det} illustrates a  typical case.  The detection depth
depends on the AO  compensation quality which was variable, reflecting
the seeing variation and airmass.   The median Strehl ratios and their
full range  are 0.36  and (0.15,  0.59) in the  red channel,  0.16 and
(0.08, 0.33) in  the blue channel. The Strehl  ratios calculated for a
filled  aperture are  multiplied here  by  1.08 to  account for  pupil
masking in  NICI.  The  median detection depth  in the red  channel is
$\Delta m =  5.3^m$ at 0\farcs27 and $\Delta m  = 7.4^m$ at 0\farcs90.
These detection  limits are only indicative  because actual detections
depend  on  companion's  location   and  on  details  of  the  speckle
structure.

Table~\ref{tab:measures} lists relative astrometry and photometry of
resolved pairs measured independently on the red and blue images.  For
well-resolved ($\rho > 0\farcs5$) companions the measure is obtained
by fitting the shifted and scaled image of the main companion. 
  For closer companions we used a blind deconvolution as described in
  THH10.  This procedure does not produce reliable results for the
faintest or closest companions near the detection limit. In such cases the
difference between measures in the red and blue channels informs us of
their quality.  The uncertain measures of 5 pairs have question marks in
the last column of Table~\ref{tab:measures}.

\section{Speckle interferometry and combined data}
\label{sec:speckle}

To improve statistics we  invoke the results of speckle interferometry
obtained with the 4m telescopes  Blanco and SOAR between 2008 and 2011
and  published  in  \citep{TMH10}  and  \citep{HTM11}.   These  papers
contain data on 57 astrometric  binaries from the FG-67pc sample, 9 of
which  were also  observed with  NICI.  There  are 99  stars combined.
Most observations were done in  the $I$ or Str\"omgren $y$ bands.  The
detection  limits  $\Delta m  (\rho)$  for  the  unresolved stars  are
published.  They  are not  as deep  as with NICI  since the  data were
obtained  at shorter  wavelength  and with  half  the telescope  size.
Nevertheless, 21 astrometric binaries  were resolved with speckle. The
speckle sample compiled {\it  a posteriori} from existing publications
can  be biased  towards more  resolved binaries.  
Note that we removed known binaries from the NICI program.

The merged AO  and speckle interferometry results on  the 99 stars are
presented in  Table~2 followed by comments on  the individual targets.
It contains  the Hipparcos number  and the rounded values  of parallax
$p_{\rm  HIP}$,  $\Delta   \mu$  and  acceleration  $\dot{\mu}$  (zero
indicates non-detection of astrometric perturbations in MK05).  If the
RV variability is found in GCS,  the amplitude in km/s is given in the
next column.  Otherwise it contains  flags {\bf C} (constant RV), {\bf
  SB} (known spectroscopic orbit) or {\bf -} (no RV data). The mass of
each star  is listed  in the  next column.  It  is estimated  from the
absolute   magnitude  in   $K$  band   using  photometry   from  2MASS
\citep{2MASS},  Hipparcos  parallax, and  the  standard relation  from
\citet{HM93}.   The light of  companions is  taken into  account where
necessary.   For  resolved pairs  we  estimate  the  mass ratio  $q  =
M_2/M_1$  from the  magnitude difference  using the  $K$-band standard
relation  (for  binaries resolved  with  NICI)  or  stellar models  of
\citet{Baraffe98} in  the appropriate color (for  speckle pairs).  The
separation  $\rho$  in  arcseconds   also  is  given.   Then  we  list
order-of-magnitude estimates of orbital  periods from the third Kepler
law $P^* =  [\rho^3 \, p_{\rm HIP}^{-3} M_1  (1 + q)]^{1/2}$, assuming
that the separation equals the  semi-major axis. The flags in the next
column show  whether the  object was observed  with NICI ({\bf  n} for
unresolved, {\bf  N} for resolved,  {\bf -} if not  observed), speckle
({\bf  s}, {\bf  S}, {\bf  -}), and if it is  listed as  resolved binary  in the
Washington Double Star Catalog (WDS) \citep[][flag {\bf W}]{WDS}.  The
remarks  in the  last  column indicate  spectroscopic and  astrometric
binaries with known  orbits (in these cases the  true period is listed
instead of  $P^*$) or binary-star  designations of known pairs  in the
WDS.

\begin{figure}[ht]
\plotone{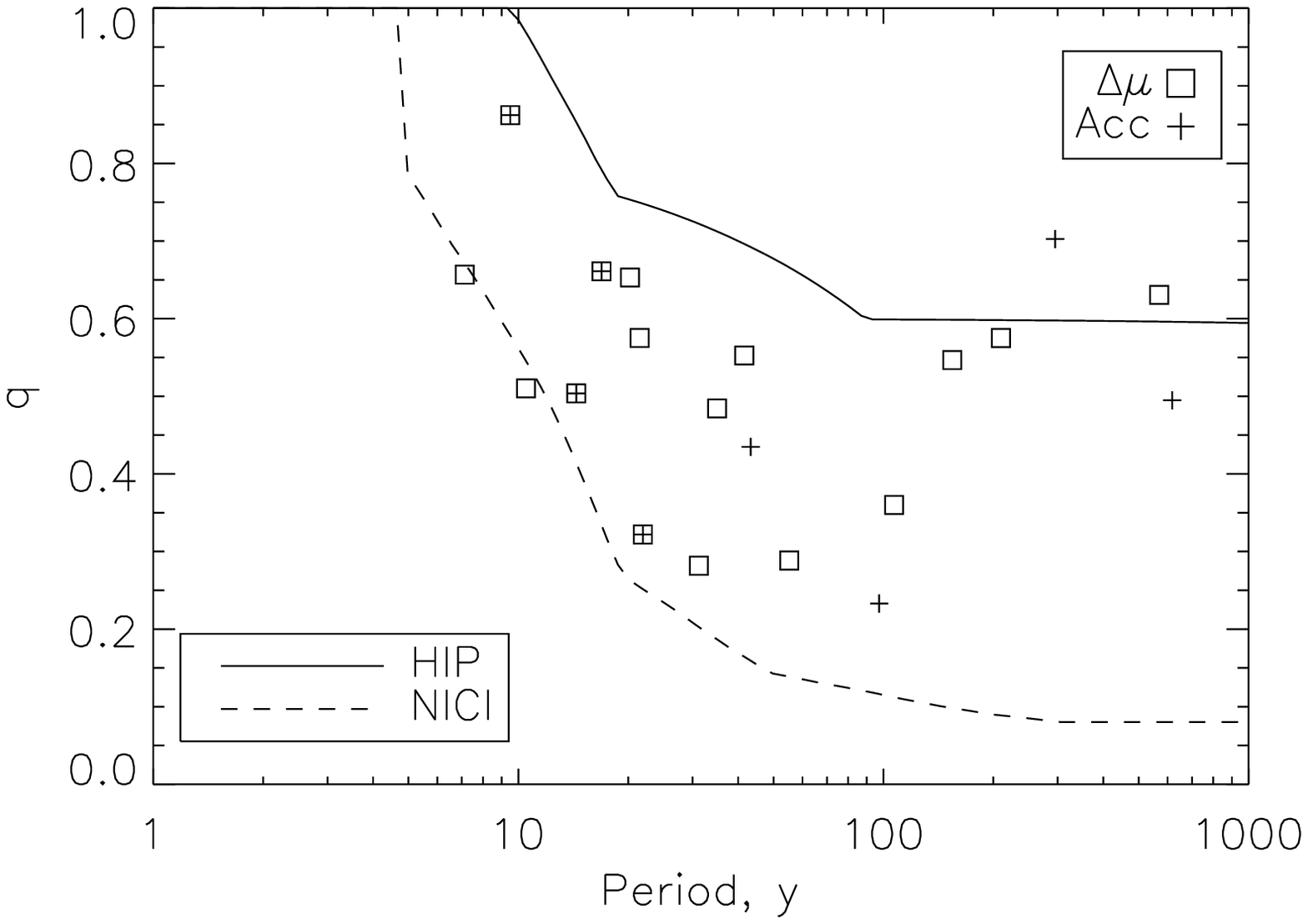}
\plotone{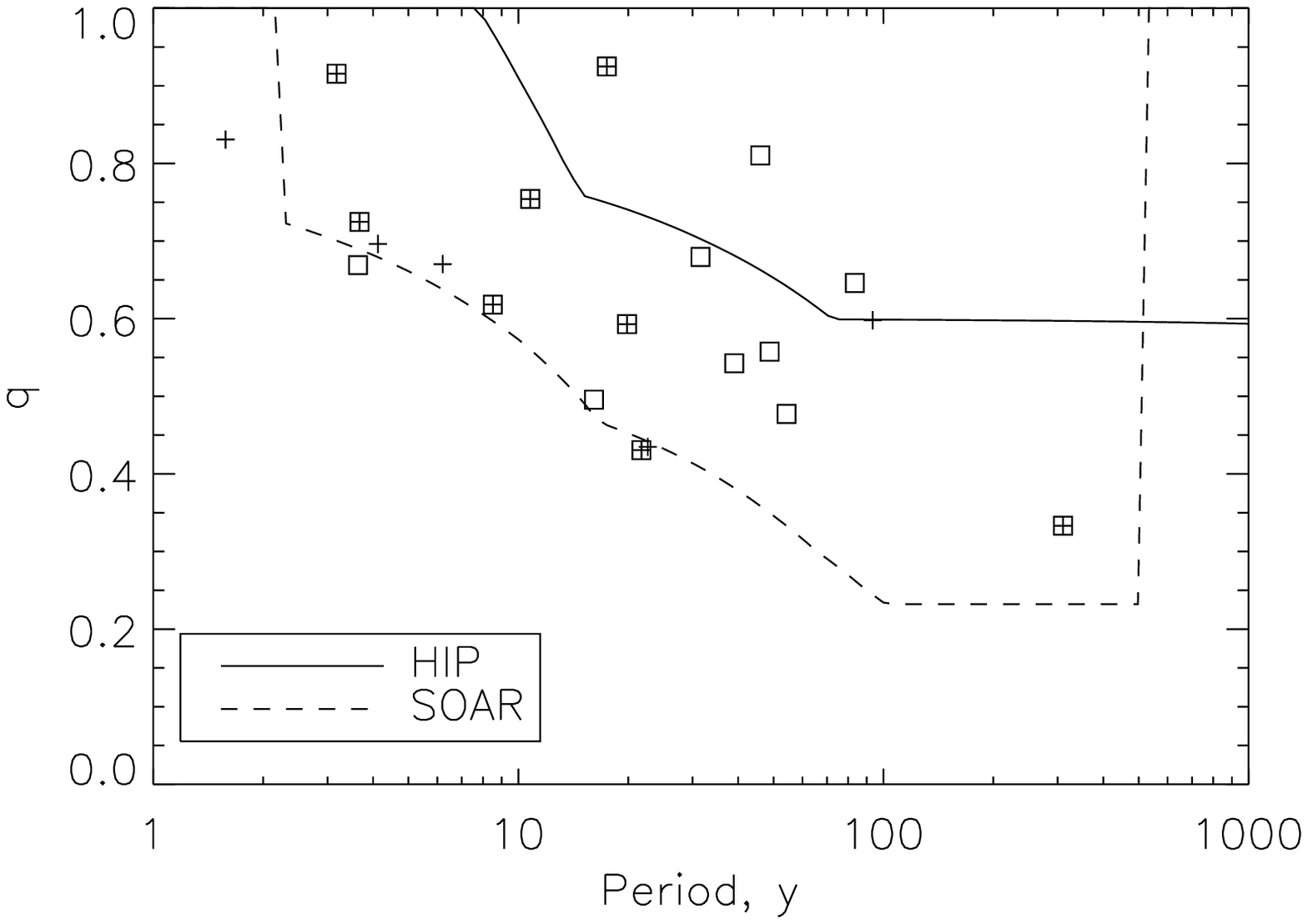}
\caption{Astrometric binaries from FG-67pc actually resolved with NICI
  (top) and speckle (bottom) in the $(P,q)$ plane.  Squares and pluses
  mark $\Delta \mu$ and  $\dot{\mu}$ binaries from MK05, respectively.
  The curves indicate approximate detection limits at 50\,pc distance.
\label{fig:plotdet} 
}
\end{figure}

The fraction  of directly resolved astrometric  binaries is 21/51=0.41
for  NICI  (3  companions  wider  that  $3''$  are  not  counted)  and
21/57=0.37 for  speckle.  These numbers  should be taken  with caution
because some  resolved companions may  belong to triple  systems where
the close inner pair  is responsible for the astrometric acceleration.
Note that our speckle data  contain a substantial number of previously
known binaries.  About 10\% of the Hipparcos astrometric binaries from
MK05 are listed in the WDS as resolved; such objects were removed from
the  NICI program.   Figure~\ref{fig:plotdet}  compares parameters  of
systems resolved  here with NICI and speckle  with estimated detection
limits.

Table~2 lists  67 spectroscopic binaries (SB) including  11 with known
orbits  (we do  not  count here two short-period  sub-systems), 13  stars
without RV data and 17 stars with constant RV.  Overall, the detection
rate of astrometric binaries by RV is high, 67/87=77\%.  However, some
of those SBs were missed by the GCS.

\citet{Frankowski}  used a  slightly different  approach  to detecting
astrometric  binaries than MK05 (total acceleration  and $\Delta \mu$
instead  of  their  components  along coordinate  axes  and a different
statistical criterion).   They do  not confirm long-term  $\Delta \mu$
acceleration in  seven systems from Table~2, namely  HIP 20524, 25180,
31480, 48072,  88648, 108649, 109067  (all but HIP~88648  have $\Delta
\mu  < 10$\,mas~y$^{-1}$).   However, HIP~108649  was resolved  with  NICI at
0\farcs812, HIP~109067 is a 4.6-y SB, and two more stars have variable
RV. Therefore we cannot affirm that their analysis is better than that
of  MK05. 

The  new   Hipparcos  reduction,  HIP2   \citep{HIP2},  contains  fewer
acceleration binaries  than the original  catalog, HIP. Of  2403 stars
with non-zero  $\dot{\mu}$ and $p_{\rm  HIP} > 15$\,mas in  MK05, half
(1244) have standard 5-parameter  solutions in HIP2 (no acceleration),
the rest are divided equally between acceleration (7- and 9-parameter)
or stochastic solutions (large residuals without polynomial or orbital
fits).  On  the other hand,  some new acceleration binaries  appear in
HIP2.  We  are not in a  position to compare  treatment of astrometric
binaries in  HIP and  HIP2 and focus  here only  on the 99  objects in
Table~2, for  which HIP2 gives  52 standard 5-parameter  solutions, 24
accelerations  (which   agree  well  with  HIP),   and  23  stochastic
solutions.  There  are 15 stars  with non-zero $\dot{\mu}$ in  HIP but
standard  5-parameter solutions in  HIP2 (the  HIP numbers  493, 5697,
8653, 12425,  13350, 17478, 24336, 25905, 28083,  38134, 45995, 49767,
103260, 112052,  114313).  Of those 15, 7  are spectroscopic binaries,
three  more are resolved  with NICI,  while other  5 stars  with small
$\dot{\mu}$   may   have    spurious   accelerations   in   HIP   (see
Sect.~\ref{sec:bogus}).

It is  perplexing that  several acceleration binaries  have companions
with separations on the order of $1''$.  Nine pairs (HIP 11072, 11537,
12145, 16853,  21543, 24336, 103260, 109443,  116125) are acceleration
binaries  without detectable  $\Delta \mu$  yet with  relatively large
separations  (hence  long  periods).   Some  of those  may  owe  their
acceleration  to unresolved inner  sub-systems, for  example HIP~21543
with a  2.2\,y inner SB  (which however should not  produce detectable
acceleration according to our simulations).

The case  of HIP~11072 ($\kappa$~For) deserves  special comment. The joint
analysis of astrometric  \citep{GK02}, spectroscopic \citep{Abt06} and
visual \citep{HTM11}  orbits, to be presented elsewhere,  leads to the
firm conclusion  that the astrometric  companion is as massive  as the
primary, while  it is $\sim$100  times fainter at  optical wavelengths
and has  a red color index $V-I  \sim 2$.  Most likely it  is a close  pair of
M-dwarfs.  This  explains  the  large  acceleration  of  19\,mas~y$^{-2}$
measured by Hipparcos and the  PM difference $\Delta \mu$ of 58\,mas~y$^{-1}$
between Hipparcos  and FK5.  The star is  not listed as  $\Delta \mu$
binary by MK05 simply because it is missed in Tycho-2.


\section{Simulations}
\label{sec:sim}

\begin{figure}[ht]
\plotone{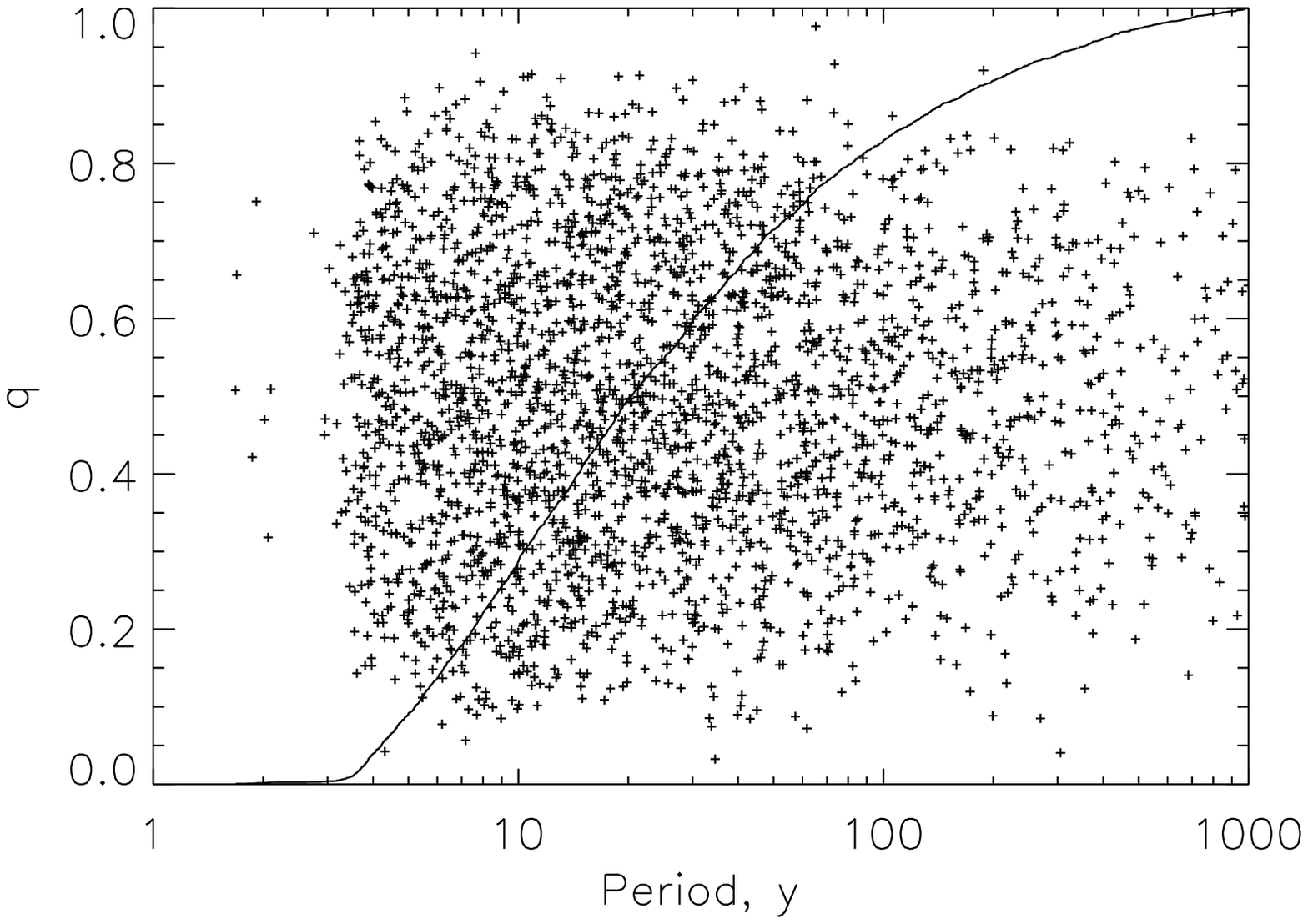}
\plotone{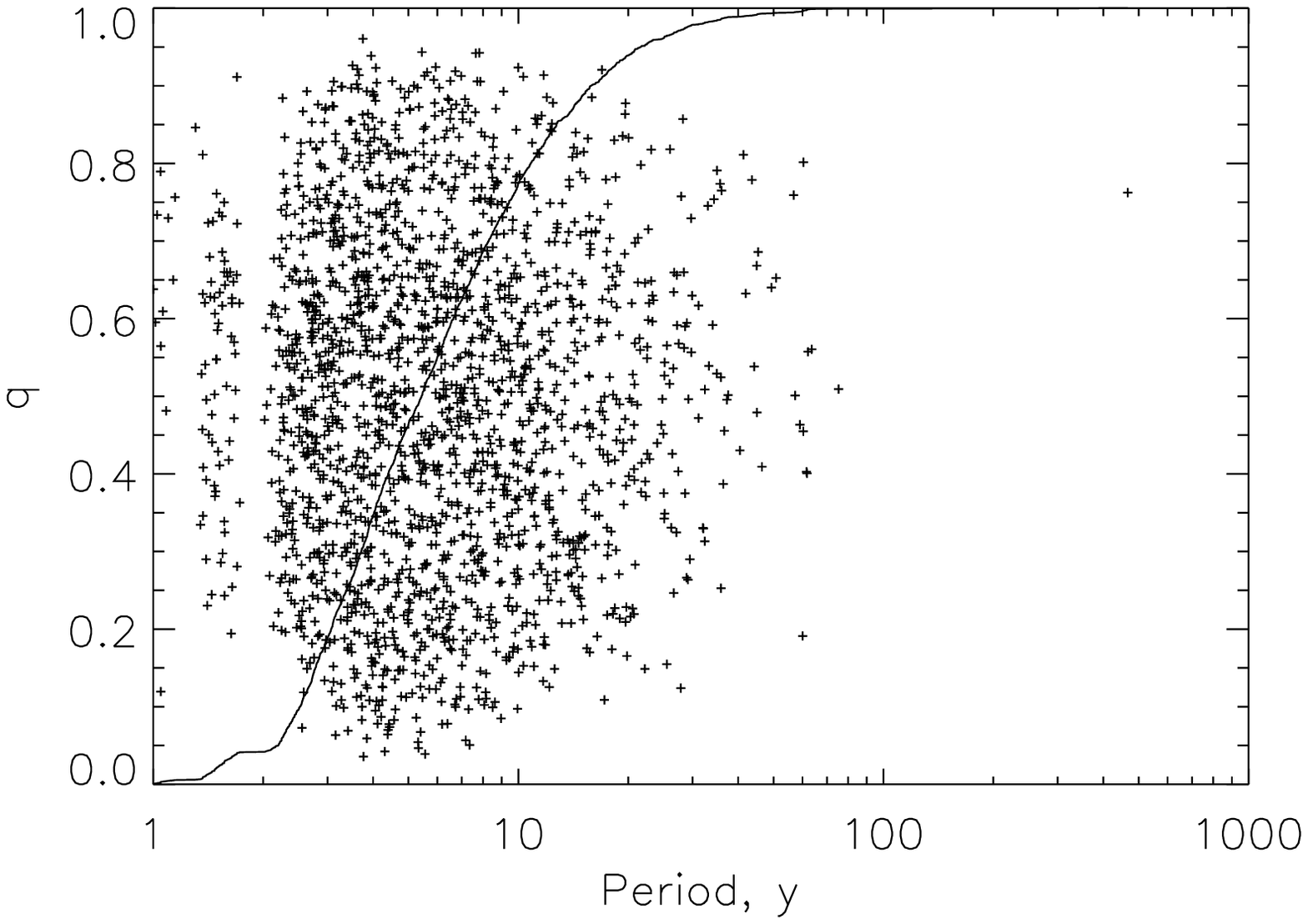}
\caption{Distributions of simulated $\Delta \mu$ (top) and $\dot{\mu}$
  (bottom) binaries in the $(P,q)$ parameter space.  The curves
  indicate cumulative period distributions. Binaries with log-uniform
  period distribution were simulated in this case. 
\label{fig:sim2} 
}
\end{figure}

To put our results in the context of binary statistics, we simulated a
large  number  of  binaries  with  dwarf components  and  $1  M_\odot$
primaries which fill  uniformly the volume up to  67\,pc (parallax $p$
larger  than 15\,mas).   Their orbital  periods are  between  1\,y and
1000\,y.  In this interval,  we use the log-normal period distribution
of  \citet{Raghavan10}  with  median  period  293\,y  and  logarithmic
dispersion  2.28.   The  mass  ratios of  companions  $q=M_2/M_1$  are
uniformly  distributed.  The  orbital inclinations  $i$ are  random in
space ($\cos i$ is uniformly  distributed), and the orbital phases are
random.  The  eccentricities $e$ are  distributed as cosine  between 0
and 1, $f(e) = (\pi/2) \cos(\pi e)$, average eccentricity 0.5.

Given the  orbital period $P$  in years  and the mass  sum $1 +  q$ in
solar  masses, we determine  the semi-major  axis $a$  in A.U.  by the
third Kepler law as $a = [P^2 (1 + q)]^{1/3}$.  The semi-major axis of
the astrometric (photo-center) orbit equals $\alpha = a p \phi$, where
the factor $\phi$ accounts for the  mass ratio $q$ and the light ratio
$r$,
\begin{equation}
\phi = \frac{\alpha}{a p} = \frac{q - r}{(r+1)(q+1)} .
\label{eq:phi}
\end{equation}
We assume $r = q^{3.75}$  -- an approximation of the standard relation
for dwarfs below $1 M_\odot$  in the $V$ band. This assumption affects
only high-$q$  binaries, for the remaining ones  the companion's light
is  negligible  and  the  photo-center  motion depends  only  on  $q$.
The maximum astrometric effect $\phi =  0.26$ is produced by binaries with
$q \sim 0.5$.

To mimic Hipparcos observations, we simulate 10 measurements uniformly
spaced  in time  $t$ from  $-T/2$ to  $T/2$, where  $T=3.2$\,y  is the
duration   of  the   Hipparcos  mission.   The  displacement   of  the
photo-center  in  $X$,  $Y$  caused  by  a motion  due  to a binary  is
calculated  for  each of  these  10 instants  in  time  and fitted  by
parabolas, for  example $X(t)  \approx a +  b t  + c t^2$.   Then, the
binary-related component of the PM is $\mu_x = b$ and the acceleration
is  $\dot{\mu}_x =  2c$  provided  that $t$  is  symmetric around  the
coordinate origin.  We identify $\sqrt{\mu_x^2 + \mu_y^2}$ with the PM
difference  $\Delta \mu$, assuming  that the  Tycho-2 PM  reflects the
true  center-of-mass  motion of  each  system.   Similarly, the  total
acceleration  is $\dot{\mu}  = \sqrt{\dot{\mu}_x^2  + \dot{\mu}_y^2}$.
Recall that  MK05 used  $\Delta \mu$ and  $\dot{\mu}$ in  each coordinate
separately     for    detecting     astrometric     binaries,    while
\citet{Frankowski}   used   total    motion   and   obtained   similar
results.

Typical   separations  of  100-y   binaries  are   around  $0\farcs3$,
comparable  to   the  grating  period   in  Hipparcos,  $1\farcs2074$.
Therefore our implicit assumption  that astrometric motion measured by
Hipparcos refers to the true  photo-center of the combined light is no
longer true  and the situation  is more complex.   Moreover, Hipparcos
measured stellar positions  in one dimension with a  scanning law that
is specific  to each star. For  these reasons the  simulations are not
an exact match to reality.

In our simulations, the binary  is considered detected by Hipparcos if
$\Delta  \mu >  5$\,mas~y$^{-1}$  and/or $\dot{\mu}  > 4$\,mas~y$^{-2}$.   These
limits  are chosen  to match  the MK05  data, as  shown  below.  Among
10\,000 simulated binaries, we  find $N_{\mu} =2905$, $N_a =1767$, and
the total $N_{\mu,a}=3614$.  The ratio $N_a /N_\mu = 0.74$ is slightly
larger than 0.61  for real astrometric binaries from  FG-67pc, but the
periods of  some real $\Delta  \mu$ binaries may be longer  than 1000\,y,
driving this ratio down a bit.
 
Figure~\ref{fig:sim2} shows the  distribution of simulated astrometric
binaries  in  the $(P,q)$  plane.   The  detection  space of  the  two
astrometric techniques is clearly  defined by these plots.  The median
period of  $\Delta \mu$ binaries is  20.4\,y, 80\% of  the periods are
between 5.2\,y  and 184\,y.  The  periods of $\dot{\mu}$  binaries are
shorter: the median is 5.4\,y,  80\% of the periods between 2.5\,y and
15.8\,y.  The gap  at $P=1.6$\,y corresponds to two  orbits during the
3.2\,y mission baseline; in this case accelerations cancel out.

\begin{figure}[ht]
\plotone{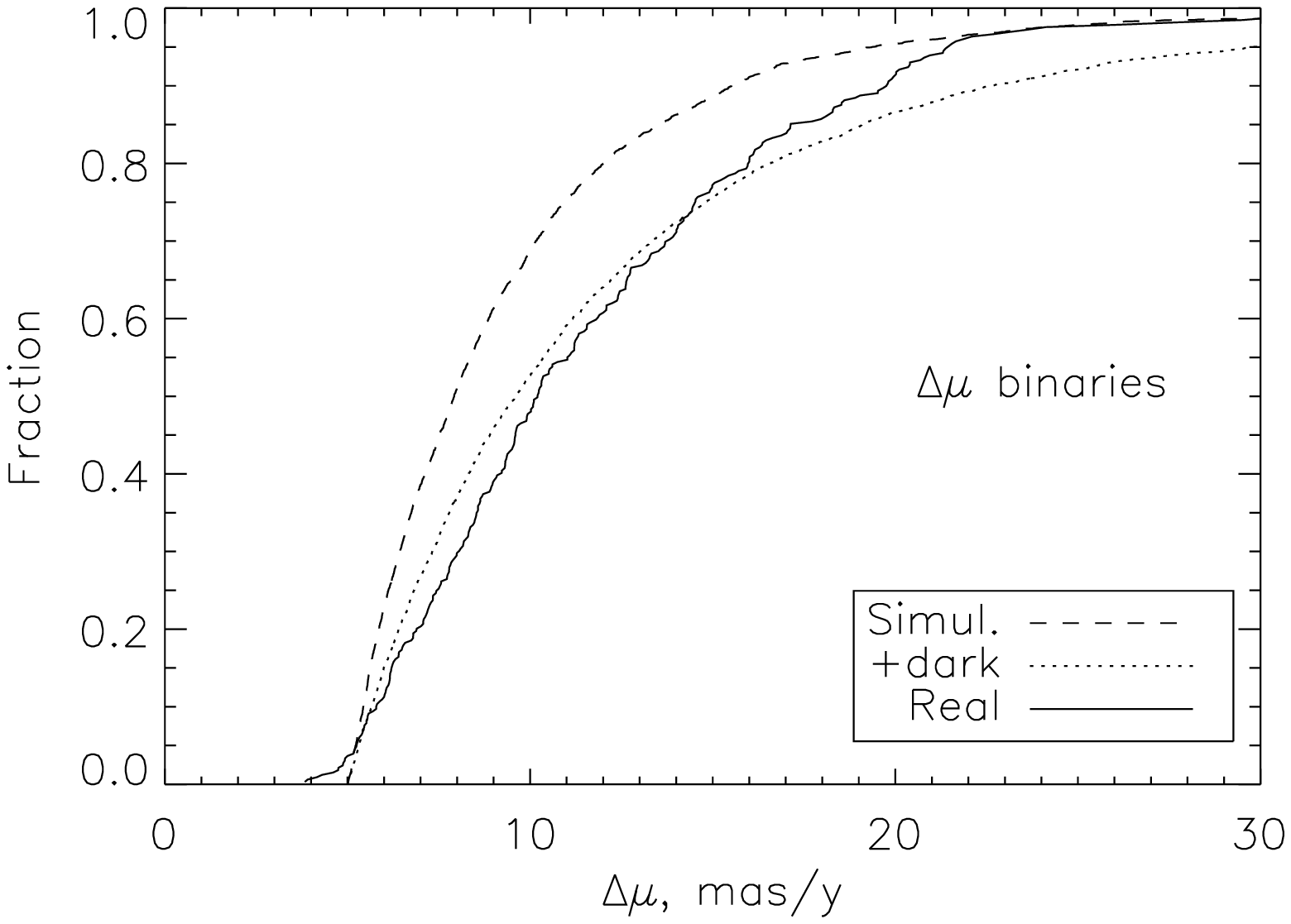}
\plotone{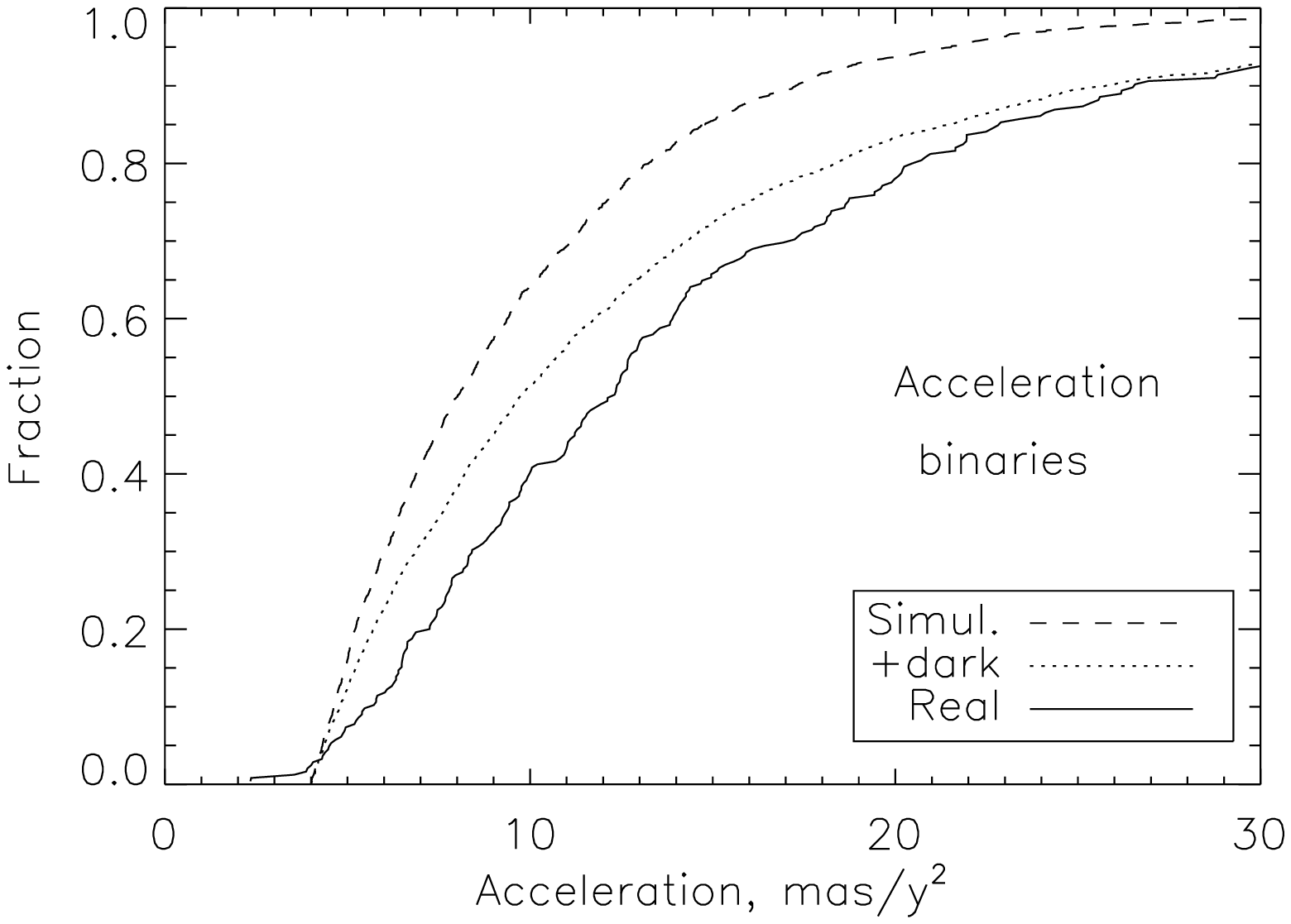}
\caption{Cumulative  distributions of  simulated and  real astrometric
  binaries.   The  simulations  are  performed  assuming  either  only
  red-dwarf companions  (dashed lines) or  a certain fraction  of dark
  and massive companions (dotted lines).  Top: distribution of $\Delta
  \mu$, bottom: distribution of $\dot{\mu}$.
\label{fig:hist} 
}
\end{figure}

We  compare  the distributions  of  $\Delta  \mu$  and $\dot{\mu}$  of
simulated and  real astrometric binaries  in Fig.~\ref{fig:hist}.  The
thresholds adopted in our simulations for the detection of astrometric
binaries by Hipparcos  are confirmed by these plots.   We see that the
real astrometric binaries from FG-67pc have, on average, the larger PM and
acceleration   compared  to  the   simulation.   The   discrepancy  in
acceleration is stronger. The discrepancy  can be reduced if we assume
that 10\%  of binaries  have white dwarf  companions with  $q=0.5$ and
additional  20\%  have dark  companions  with  $q=1$,  similar to  the
companion of HIP~11072 (dotted  lines in Fig.~\ref{fig:hist}). In this
case  the simulated  fraction  of acceleration  binaries $N_a/N_\mu  =
0.56$   also  becomes   closer  to  the 0.61 fraction in MK05. 
We  should bear  in  mind that  our  simulations involve  a number  of
simplifying assumptions and that  the observed parameters $\Delta \mu$
and $\dot{\mu}$  can be  affected by errors.  Therefore the  degree of
agreement with our simulations is quite satisfactory.

\begin{figure}[ht]
\plotone{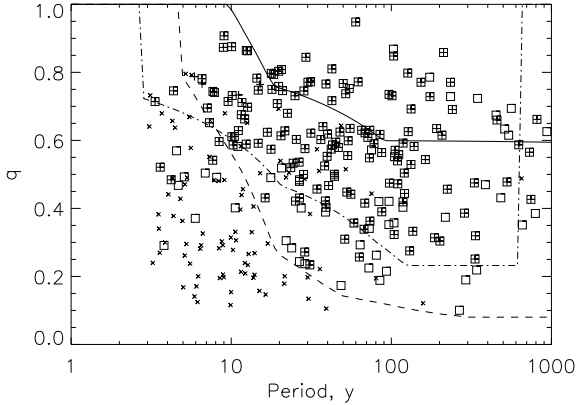}
\caption{Simulated  direct resolution of  353 astrometric  binaries by
  NICI  (squares, $N=195$) and  speckle (pluses,  $N=151$). Unresolved
  objects  are   plotted  as   small  crosses.  The   curves  indicate
  approximate limits of direct  resolution with Hipparcos (full line),
  NICI  (dashed  line)  and  speckle (dash-dot line)  for  binaries  at
  50\,pc.
\label{fig:simdet} 
}
\end{figure}

For evaluating the fraction of objects resolved with NICI and speckle,
we simulated 1000 binaries with  the same statistics as above (without
dark companions), of which 337 turned out to be detectable astrometric
binaries.   Their separations  at a random  moment of time  and magnitude
differences in the $V$ (Hipparcos), $I$ (speckle) and $K$ (NICI) bands
are  compared to the  detection limits  of respective  techniques, and
objects    above    those    limits   are    declared    ``resolved''.
Table~\ref{tab:detlim}  lists  the  adopted detection  limits  $(\rho,
\Delta m)$,  interpolated linearly between these points.   We find 195
binaries ``resolved''  with NICI  and 151 with  speckle (147  of those
also with  NICI).  The fraction  of simulated resolutions with  AO and
speckle  is  0.55  and  0.43  respectively,  higher  than  the  actual
fractions of 0.41 and  0.37.  The separation distributions of actually
resolved binaries and simulated  resolved binaries match well both for
NICI and for  speckle, in all cases the median  separations are in the
interval between 0\farcs27 and 0\farcs30.

The fraction of binaries resolved with NICI can be reconciled with our
simulations if  we assume that there  are $\sim 5$  dark companions in
the observed sample  of 51 stars (subtract from  the total number) and
take  into  account  the  removal of  $\sim$10\%  previously  resolved
systems from the NICI program (add 5 to the total and resolved count).
The  corrected resolution  rate  is then  $(21 +  5)/(51  + 5  - 5)  =
0.51$. If we consider only $\Delta \mu$ systems, the resolved fraction
16/34=0.47  is still  less than  the 0.68 predicted  by  simulation.  The
agreement  can  be  restored  to  within  statistical  uncertainty  by
assuming $\sim$10\% of dark companions.

\section{Spurious accelerations in Hipparcos}
\label{sec:bogus}

A subset of Hipparcos stars marked with the acceleration flag ``G'' in
the catalog has drawn special attention because of their apparently
enigmatic nature.  These stars have small parallaxes, sometimes even
unphysical negative values, indicating large distances, and
yet, their large accelerations suggest orbits of considerable size.
Using the simple formulae from \citep{kama} or the Q-factor from MK05,
we obtain incongruously large lower bounds for the masses of the
invisible companions.  Although multiple systems with dim companions
which comprise a tight binary with a total mass exceeding the mass of
the primary do exist, the lower limit masses of the most distant
accelerating stars suggest the presence of more exotic or hypothetical
objects, such as stellar-mass black holes or failed supernova
\citep{gould}.

Table~\ref{tab:acc} includes a sample  of some of the most interesting
distant  accelerating  stars   from  the  original  Hipparcos  catalog
\citep{HIP1997d} with parallaxes $<2$\,mas (none of them belong to our
FG-67pc  sample). The large  distances for  these stars  inferred from
Hipparcos parallaxes are confirmed by secondary criteria. The star HIP
111759  has  been observed  after  the  publication  of the  Hipparcos
catalog spectroscopically (J.   Sperauskas, private communication) and
by  ground-based interferometry,  providing no  clue of  binarity. The
negative  result  for  this  star  casts  doubt  on  the  validity  of
accelerations for the entire sample.  However, the formal significance
of the accelerations, given in the last column of the table, is always
greater than 3.5$\sigma$, implying  a very small probability of error.
Even more confusingly, binarity of these stars appears to be confirmed
by significant $\Delta \mu$  (columns 3--4).  

We performed further computations for the stars in Table~\ref{tab:acc}
using  the  HIAD.   For  each  star,  we  solved  the  entire  set  of
observational equations  by a  weighted least-squares method,  for the
standard set of five astrometric  unknowns.  In each case, the results
for proper  motion components  {\it without} solving  for acceleration
components turned out to  be statistically consistent with the Tycho-2
proper  motions.   Thus, the  astrometric  solutions  in the  original
Hipparcos  catalog  were  strongly   perturbed  by  the  inclusion  of
additional acceleration unknowns in the observational equations.  This
gives a  clear clue that the  accelerations for the 11  stars in Table
\ref{tab:acc}   are  spurious.      Comparison   between   5-  and
  7-parameter solutions  for HIP~2777  shows that the  acceleration is
  not statistically significant, resulting from the poorly-conditioned
  system  of  least-squares equations.   The  new Hipparcos  reduction
  \citep{HIP2}   gives  5-parameter   solutions  for   all   stars  in
  Table~\ref{tab:acc} (except for one stochastic solution i.e. HIP~107466);
  the HIP2 PMs, also listed in Table~\ref{tab:acc}, are close to those
  of Tycho-2.

\section{Discussion}
\label{sec:disc}

Our  direct  imaging  resolved  about  40\%  of  Hipparcos  astrometric
companions. Although at  least half of those were  also detected by RV
variability, their  estimated periods exceed  10\,y and only  a few have
known SB orbits.  In contrast, imaging  is a fast and efficient way to
estimate periods and mass ratios of resolved binaries.

Our simulations show that our results are in reasonable agreement with
statistics  of  solar-type  binaries within  25\,pc  \citep[log-normal
  period    distribution   and   uniform    mass-ratio   distribution,
  see][]{Raghavan10}.    The  separations   of   resolved  pairs   are
distributed in agreement with  these simulations, with a median around
0\farcs3.  The  fraction of  directly resolved companions  is slightly
less than  expected.   This can be  explained by the  presence of
  $\sim$10\%  of ``dark'' companions  -- white  dwarfs (WD)  and close
  pairs of M-dwarfs.  The  expected fraction of WDs (former primaries)
  depends on such  unknowns as multiplicity of massive  stars and star
  formation  history.   \citet{DM91}  quote  a rough  estimate  of  WD
  companions  1.2\% per  decade  of period  in  their Sect.~6.2.   The
  fraction of  secondary companions that are themselves  binary is not
  constrained observationally, given  the difficulty of detecting such
  sub-systems. In the  simulations we adopted 10\% of  WDs and 20\% of
  binary secondaries to illustrate the effect of massive companions on
  the distribution  of acceleration  and $\Delta \mu$.   However these
  distributions   are  influenced   by  several   other   factors  and
  assumptions,  therefore  the  improved   fit  in  Fig.~5  cannot  be
  considered as a safe estimate of dark-companion frequency.

We believe  that the majority  of unresolved astrometric  binaries are
real, but their  companions are just too faint and/or  too close to be
resolved.  In few cases this is confirmed by SB orbits.  Yet, our work
reveals some problems and unsolved questions.

First, we  find that  9 acceleration-only binaries  have in  fact wide
companions,  too wide  to  cause the  acceleration.  Considering  that
Hipparcos  measured stellar positions  by 1-dimensional  scanning with
grating,  it  is  plausible  that faint  companions  with  separations
comparable to the grating  period 1\farcs2074 caused systematic errors
that mimic acceleration.  The issue could be further  studied 
by modeling, although a companion with  $\Delta m = 5$ produces only a
small effect of few mas.

Second, we  show that  the original Hipparcos catalog  contains some
spurious accelerations  accompanied by  erroneous PMs. This  is proven
for  distant  stars   whose  accelerations  imply  improbably  massive
invisible  companions, should  they  be real.  Nearby  stars may  also
suffer from  such errors caused by  fitting too many  parameters in an
ill-conditioned  least-squares   problem.   The  new   reduction  HIP2
contains less stars with  acceleration,  eliminating some spurious
  acceleration solutions but also  missing some real astrometric binaries.
The  situation  is  not clear  and  has  to  be addressed  by  careful
examination of  each individual case.  For  now, acceleration binaries
 in HIP and HIP2  should be considered with caution and confirmed
by other techniques whenever possible.

Simulations  demonstrate  that   Hipparcos  astrometry  combined  with
Tycho-2  readily  detects  stellar-mass  companions to  dwarfs  within
67\,pc  with  periods from  few  to  few  hundred years.   Sub-stellar
companions (brown dwarfs) can  be discovered only in exceptional cases
for the nearby stars;  in  addition, such  companions are  intrinsically
rare  (brown dwarf dessert).

The  situation  will  change  dramatically  when  high-precision  GAIA
astrometry becomes available. Owing  to the short 5-y mission duration
and the lack of an  accurate long-term reference analogous to Tycho-2,
only the $\dot{\mu}$ method of binary detection will be valid; it will
be able to reveal companions of planetary mass.  However, some lessons
learned from Hipparcos astrometric  binaries will be still relevant at
this new level of precision.


\acknowledgments  We  thank  Fredrik  Rantakyro (Gemini)  for  careful
observations  of  our  program  stars.  The comments  by  the  referee
A.~Sozzetti  have been  much  appreciated.  This  work  used the  SIMBAD
service  operated  by  Centre  des Donn\'ees  Stellaires  (Strasbourg,
France),  bibliographic references from  the Astrophysics  Data System
maintained by SAO/NASA, data products of the Two Micron All-Sky Survey
(2MASS) and the Washington Double Star Catalog maintained at USNO.
Gemini telescopes  are  operated by the  Association of Universities
for Research  in Astronomy, Inc.,  under a cooperative  agreement with
the  NSF on  behalf of  the Gemini  partnership: the  National Science
Foundation  (United  States), the  Science  and Technology  Facilities
Council  (United  Kingdom), the  National  Research Council  (Canada),
CONICYT   (Chile),  the   Australian  Research   Council  (Australia),
Minist\'{e}rio da Ci\^{e}ncia e  Tecnologia (Brazil) and Ministerio de
Ciencia, Tecnolog\'{i}a e Innovaci\'{o}n Productiva (Argentina).


{\it Facilities:} \facility{Gemini:South (NICI)}

\clearpage

\begin{deluxetable}{l c rcc rcc l}            
\tabletypesize{\scriptsize}                                                                                                               
\tablecaption{Measures of  companions resolved with NICI
\label{tab:measures}   }                                                                                                                       
\tablewidth{0pt}                                                                                                                          
\tablehead{  HIP   & Date & \multicolumn{3}{c}{Red 2.272\,$\mu$m} &  \multicolumn{3}{c}{Blue 1.587\,$\mu$m} &  Rem \\ 
                   &  &   P.A.         &  Sep.   & $\Delta m$ &     P.A.         &  Sep.   & $\Delta m$ &  }
\startdata                                                                                                       
5144 & 2011.8433 &       38.6 &    0.254 &     3.96 &       41.0 &    0.251 &     4.46 &  \\
6273 & 2011.8434 &      117.0 &    0.269 &     2.29 &      117.5 &    0.270 &     2.79 &  \\
6712 & 2011.8434 &        9.2 &    0.104 &     0.64 &        6.4 &    0.094 &     0.72 &  \\
8498 & 2011.8490 &      283.0 &    0.272 &     2.46 &      282.9 &    0.272 &     2.67 &  \\
11537 & 2011.8490 &      347.6 &    4.113 &     3.48 &      347.7 &    4.119 &     3.92 &  \\
12425 & 2011.8383 &       77.9 &    0.344 &     3.77 &       80.9 &    0.334 &     4.51 & ? \\
14527 & 2011.7753 &      194.8 &    0.295 &     3.65 &      193.8 &    0.289 &     3.50 &  \\
16853 & 2011.7753 &       90.0 &    2.696 &     3.26 &       90.0 &    2.683 &     3.69 &  \\
21079 & 2011.7752 &       23.4 &    1.622 &     1.91 &       23.4 &    1.619 &     2.14 &  \\
21778 & 2011.8027 &      176.2 &    0.165 &     3.03 &      177.2 &    0.151 &     3.19 & ? \\
22387 & 2011.7752 &       64.5 &    0.165 &     3.76 &       68.1 &    0.158 &     3.96 & ? \\
24336 & 2011.7972 &      351.3 &    1.250 &     1.46 &      351.4 &    1.246 &     1.64 &  \\
25148 & 2011.8027 &      195.2 &    0.054 &     1.29 &      188.5 &    0.065 &     1.11 & ? \\
28241 & 2011.6960 &      356.4 &    0.539 &     2.50 &      355.5 &    0.542 &     2.79 &  \\
97676 & 2011.6978 &       33.6 &    0.156 &     1.91 &       34.2 &    0.156 &     2.15 &  \\
103260 & 2011.6979 &      359.0 &    3.975 &     1.83 &      359.0 &    3.968 &     2.08 &  \\
108041 & 2011.8512 &      114.1 &    0.812 &     2.29 &      113.8 &    0.814 &     2.49 &  \\
108649 & 2011.8513 &       39.6 &    0.202 &     2.94 &       40.2 &    0.199 &     3.33 &  \\
109443 & 2011.8513 &      346.3 &    1.420 &     2.92 &      346.2 &    1.433 &     3.18 &  \\
114880 & 2011.8512 &      146.8 &    0.100 &     2.52 &      146.7 &    0.090 &     3.27 & ? \\
115505 & 2011.8512 &      262.6 &    0.462 &     3.65 &      262.5 &    0.461 &     4.20 & AB \\
115505 & 2011.8512 &      319.4 &   13.735 &     4.61 &      319.7 &   13.778 &     5.47 & AC \\
116125 & 2011.8512 &      143.1 &    0.239 &     3.45 &      143.5 &    0.239 &     3.73 &  \\
117258 & 2011.8513 &       18.6 &    0.228 &     1.76 &       18.9 &    0.227 &     2.13 &  \\
\enddata                                                                                                                       
\end{deluxetable}

\begin{deluxetable}{l ccc cc ccc l l }            
\tabletypesize{\scriptsize}                                                                                                               
\tablecaption{Summary data on observed astrometric binaries
\label{tab:all}   }                                                                                                                       
\tablewidth{0pt}                                                                                                                          
\tablehead{  HIP   & $p_{HIP}$ & $\Delta \mu$ & $\dot{\mu}$ & $\Delta$RV &$M_1$ & $q$ & $\rho$ & $P^*$ & Flags & Remark \\
                   & mas & mas~y$^{-1}$        & mas~y$^{-2}$ & km~s$^{-1}$ & $M_\odot$ &    & arcsec & y   &  N S W     & }
\startdata    
   493 &   26 &    0 &    4 &     C  &   0.97 &        &        &        & - s -  &  \\
  1573 &   22 &   13 &    0 &    1.3 &   1.12 &        &        &        & - s -  &  \\
  2066 &   16 &   12 &    0 &   -    &   1.05 &   0.65 &  0.386 &   28.8 & - S W  & YR 4 \\
  5144 &   23 &    8 &    0 &    C   &   1.04 &   0.28 &  0.254 &   31.2 & N - -  &  \\
  5697 &   19 &   18 &   14 &    SB  &   0.86 &        &        &   39.7 & n - -  & SB1, $q>$0.2 \\
  6273 &   30 &   19 &    0 &    2.2 &   0.92 &   0.58 &  0.269 &    8.9 & N - -  & AB \\
  6712 &   18 &   16 &   25 &    0.8 &   0.93 &   0.86 &  0.102 &    9.5 & N - -  &  \\
  7142 &   15 &    6 &    0 &   -    &   0.84 &        &        &        & n - -  &  \\
  7580 &   24 &   19 &    8 &    SB  &   1.34 &   0.72 &  0.079 &   28.8 & - S W  & SB2, KUI 7 \\
  7869 &   15 &    0 &   12 &    6.9 &   1.09 &        &        &        & n s -  & sb2 \\
  8498 &   19 &    9 &    0 &    1.2 &   1.06 &   0.55 &  0.273 &   41.6 & N S -  &  \\
  8653 &   21 &    0 &    2 &   -    &   0.85 &        &        &        & n - -  &  \\
 10611 &   16 &   10 &    7 &   -    &   0.96 &   0.92 &  0.044 &    3.2 & - S -  &  \\
 11072 &   45 &    0 &   19 &    5.1 &   1.28 &   0.43 &  0.465 &   26.5 & - S W  & LAF 27, VB \\
 11537 &   16 &    0 &   26 &   -    &   0.92 &        &        &        & N - -  & Comp. at 4.1" \\
 12425 &   15 &    0 &   17 &    C   &   0.98 &   0.23 &  0.359 &   97.3 & N - -  &  \\
 12716 &   24 &    9 &    0 &    SB  &   1.03 &   0.56 &  0.378 &   48.8 & - S -  & SB2 1d, triple \\
 12843 &   70 &    0 &   26 &    C   &   1.15 &        &        &        & - s -  &  \\
 12889 &   20 &   10 &   10 &    1.8 &   1.09 &        &        &        & n s -  &  \\
 13350 &   17 &   10 &    6 &    C   &   0.88 &        &        &        & n - -  &  \\
 14527 &   19 &    8 &    0 &    1.5 &   0.95 &   0.29 &  0.297 &   55.2 & N s -  &  \\
 16370 &   20 &    0 &   18 &    1.4 &   1.27 &   0.70 &  0.070 &    4.1 & - S -  & sb2 \\
 16851 &   19 &    6 &   11 &    2.0 &   1.26 &        &        &        & n - -  &  \\
 16853 &   23 &    0 &    5 &    4.0 &   1.02 &   0.41 &  2.696 & 1056.2 & N s -  & triple? \\
 17108 &   16 &    0 &    6 &    1.6 &   1.36 &        &        &        & n - -  &  \\
 17478 &   19 &    0 &   13 &    2.5 &   0.99 &        &        &        & n - -  &  \\
 17820 &   20 &    0 &   30 &    4.7 &   0.94 &        &        &        & n s -  &  \\
 17895 &   19 &   14 &    0 &    9.5 &   1.23 &   0.81 &  0.329 &   46.0 & - S W  & YR 23, sb2, triple \\
 19147 &   15 &   10 &    9 &    C   &   1.14 &        &        &        & n - -  &  \\
 20375 &   18 &   13 &    0 &    1.7 &   1.13 &        &        &        & n s -  &  \\
 20524 &   20 &    5 &    0 &    C   &   0.94 &        &        &        & n - -  &  \\
 21008 &   21 &    5 &    0 &   -    &   1.15 &   0.50 &  0.170 &   16.1 & - S W  & PAT 10 \\
 21053 &   24 &    6 &    0 &    C   &   1.21 &        &  0.300 &   37.7 & - s W  & PAT 11 \\
 21079 &   20 &    5 &    0 &    3.8 &   0.94 &   0.63 &  1.622 &  569.0 & N - -  & triple? \\
 21543 &   20 &    0 &   17 &    SB  &   1.15 &   0.60 &  0.513 &   93.5 & - S W  & SB1 1.8y, CHR 153 \\
 21778 &   23 &   15 &   11 &    1.6 &   0.98 &   0.50 &  0.159 &   14.4 & N s -  &  \\
 22221 &   25 &   20 &   18 &    3.0 &   1.06 &        &  0.100 &    7.3 & - s -  & PAT 16 \\
 22387 &   18 &    9 &    8 &    2.3 &   1.05 &   0.32 &  0.165 &   21.9 & N s -  &  \\
 23818 &   26 &    0 &   19 &    C   &   1.30 &   0.83 &  0.046 &    1.6 & - S W  & FIN 376, sb2 \\
 24336 &   24 &    0 &   17 &    C   &   0.95 &   0.70 &  1.250 &  295.3 & N - -  &  \\
 24419 &   34 &    0 &   16 &    SB  &   0.93 &        &        &    2.2 & - s -  & SB1 $q>$0.14 \\
 25148 &   15 &    5 &    0 &    3.7 &   1.00 &   0.66 &  0.066 &    7.1 & N - -  &  \\
 25180 &   20 &    3 &    0 &   31.7 &   1.40 &        &        &        & - s -  &  \\
 25905 &   25 &   14 &   12 &    0.6 &   0.86 &        &        &        & n - -  &  \\
 27260 &   16 &    8 &    0 &   -    &   0.97 &        &        &        & n - -  &  \\
 27371 &   22 &   22 &   15 &    0.8 &   0.86 &        &        &        & n - -  &  \\
 28083 &   16 &    0 &    7 &   -    &   1.02 &        &        &        & n - -  &  \\
 28241 &   16 &   10 &    0 &    C   &   1.03 &   0.55 &  0.539 &  154.3 & N - -  &  \\
 28333 &   16 &    7 &   20 &    4.1 &   1.00 &        &        &        & n - -  &  \\
 29860 &   51 &    8 &    0 &     SB &   1.07 &   0.48 &  0.884 &   26.8 & - S W  & SB1, CAT 1 \\
 30480 &   31 &    6 &    0 &    1.4 &   1.17 &        &        &        & - s -  &  \\
 31480 &   15 &    7 &    0 &    4.9 &   1.28 &        &        &        & - s -  &  \\
 32329 &   20 &    5 &    0 &    6.8 &   1.19 &        &        &        & - s -  &  \\
 35642 &   27 &   11 &   12 &    1.0 &   1.18 &        &        &        & - s -  &  \\
 36836 &   22 &   14 &    0 &    3.0 &   1.21 &        &        &        & - s -  &  \\
 37853 &   65 &   64 &    6 &    1.8 &   1.06 &   0.62 &  0.326 &    8.5 & - S -  &  \\
 38134 &   19 &   11 &    6 &    5.6 &   1.05 &        &        &        & - s -  &  \\
 42408 &   24 &   14 &   17 &    1.2 &   0.92 &        &        &        & - s -  &  \\
 43299 &   24 &   20 &    0 &    2.5 &   1.04 &        &        &        & - s -  &  \\
 44874 &   21 &   18 &    0 &    2.9 &   1.31 &        &  1.700 &  503.3 & - s W  & RST 2610 \\
 44896 &   23 &    8 &    0 &    0.5 &   1.16 &        &        &        & - s -  &  \\
 45705 &   20 &   20 &   20 &    1.6 &   1.03 &   0.75 &  0.123 &   10.8 & - S W  & CHR 239 \\
 45995 &   26 &    0 &    4 &    7.5 &   1.02 &        &        &        & - s -  &  \\
 48072 &   26 &    7 &    0 &    5.0 &   1.03 &        &        &        & - s -  &  \\
 48095 &   21 &   16 &    4 &    2.6 &   1.21 &        &        &        & - s -  &  \\
 49767 &   23 &    6 &    4 &    0.8 &   1.03 &        &        &        & - s -  &  \\
 50870 &   18 &   10 &    8 &    1.6 &   1.41 &   0.33 &  1.076 &  310.2 & - S -  &  \\
 53217 &   19 &    6 &   33 &    SB  &   1.32 &        &        &    3.3 & - s -  & SB 6.8d+3.3y \\
 53424 &   19 &    0 &    9 &    8.5 &   1.00 &        &        &        & - s -  &  \\
 55714 &   21 &    8 &    0 &    1.8 &   1.01 &        &  0.100 &    7.2 & - s W  & CHR 242 \\
 59926 &   25 &   11 &   12 &    0.7 &   1.16 &        &        &        & - s -  &  \\
 60024 &   22 &   11 &    6 &    1.3 &   0.90 &        &        &        & - s -  &  \\
 64219 &   36 &   20 &   10 &    SB  &   0.92 &        &  0.310 &   23.5 & - s W  & SB1 20.4d, TOK 28 \\
 67620 &   51 &   46 &    0 &    SB  &   0.96 &   0.67 &  0.143 &   10.3 & - S W  & SB1, WSI 77 \\
 73241 &   41 &    9 &    6 &    SB  &   1.09 &   0.59 &  0.363 &   14.9 & - S W  & SB1, WSI 80 \\
 82621 &   37 &   14 &    9 &   -    &   1.22 &   0.43 &  0.359 &   21.7 & - S W  & WSI 86 \\
 85141 &   17 &   16 &    9 &    1.8 &   1.16 &   0.93 &  0.144 &   15.0 & - S W  & RST 3972, VB, sb2 \\
 88648 &   17 &   19 &    0 &   -    &   0.69 &        &        &        & n - -  &  \\
 92103 &   15 &    6 &    0 &    C   &   1.45 &        &        &        & n - -  &  \\
 97676 &   18 &   23 &   31 &    2.2 &   0.95 &   0.66 &  0.145 &   16.9 & N - -  &  \\
103260 &   22 &    0 &    7 &    C   &   1.11 &   0.64 &  3.975 & 1694.7 & N - W  & I 18 \\
103735 &   21 &   10 &    0 &    3.4 &   1.09 &        &        &        & n - -  &  \\
103987 &   19 &    0 &   21 &    SB  &   1.23 &   0.67 &  0.084 &    1.0 & - S W  & SB1, WSI 6 \\
108041 &   20 &   15 &    0 &    C   &   0.88 &   0.57 &  0.812 &  209.8 & N - -  &  \\
108095 &   18 &    8 &    0 &    C   &   0.91 &        &        &        & n - -  &  \\
108649 &   15 &    5 &    0 &    C   &   1.14 &   0.48 &  0.199 &   35.0 & N - -  &  \\
109067 &   19 &    7 &    0 &    SB  &   0.76 &        &        &    4.6 & n - -  & SB1 $q>$0.23 \\
109443 &   16 &    0 &   11 &    2.5 &   1.16 &   0.49 &  1.420 &  617.7 & N - -  &  \\
109470 &   16 &    0 &   21 &    2.2 &   1.26 &        &        &        & n - -  &  \\
110649 &   48 &   33 &    0 &    1.5 &   1.21 &        &        &        & - s -  &  \\
112052 &   18 &    0 &   16 &   -    &   0.85 &        &        &        & n - -  &  \\
112506 &   25 &   26 &    0 &   -    &   0.96 &   0.68 &  0.304 &   31.5 & - S W  & WSI 93 \\
114313 &   15 &    0 &   37 &    SB  &   0.88 &        &        &    3.1 & n - -  & SB1 \\
114880 &   16 &    8 &    0 &    2.5 &   1.05 &   0.51 &  0.095 &   10.5 & N - -  &  \\
115505 &   17 &    9 &    0 &    2.6 &   1.09 &   0.36 &  0.462 &  107.0 & N - -  & triple? \\
116125 &   15 &    0 &   21 &    C   &   1.25 &   0.43 &  0.239 &   43.3 & N - -  &  \\
117258 &   25 &   11 &    0 &    1.4 &   1.01 &   0.65 &  0.228 &   20.2 & N - -  &  \\
117493 &   15 &   15 &    4 &    4.1 &   1.37 &        &        &        & n - -  &  \\
117513 &   15 &    0 &   13 &   -    &   1.01 &        &        &        & n - -  &  \\
\enddata                                                                                                                       
\end{deluxetable}

\clearpage
\centerline{\bf Notes to Table 2}

\begin{scriptsize}
HIP 493: The visual binary HIP~495AB at 573$''$ is co-moving, same parallax.
         HIP 493 is $\sim 1.3^m$ below the Main Sequence in the $(K,V-K)$ color-magnitude diagram.

HIP   5697:  The SB1   orbit  suggests   mass  ratio   $q>0.2$   and  axis
0\farcs21. The companion is probably just a bit too faint or too close
to be resolved with NICI.

HIP  6273: Two  astrometric orbits  are derived  by  \citet{GM06}, the
          longest one has $P=8.93$y, axis 55\,mas, $e=0.84$.

HIP 6712: Possible physical companion at 24\farcs4, 227$^\circ$ in 2MASS.

HIP 8653: Small acceleration of 2 mas/y$^2$, possibly single.

HIP  11072: The astrometric binary with 26.5-y orbit has a massive
     secondary with $q \sim 1$ (see text). 

HIP 11537: The 4$''$ companion  is partially resolved in 2MASS images,
           for  that  reason  the  star  is not  found  in  the  2MASS
           point-source catalog. The companion is physical, because it
           keeps   the  same   position  for   10\,y  despite   PM  of
           0\farcs2~y$^{-1}$.   This  is   a   pre-Main-Sequence  star
           according to SIMBAD.
    
HIP 12425: A faint companion near the detection limit is found, our measures
     are uncertain.  The estimated  long period $P \sim 100$y suggests
     that Hipparcos aceleration of 17\,mas~y$^{-2}$ is spurious.

HIP 12716:  Triple.  The  A-component is 0.955d  SB2, the  tertiary is
resolved with  speckle.  As A is  located some 1.7$^m$  above the Main
Sequence,  the  object  can  be closer  than  indicated  by  the
Hipparcos parallax of 24.6\,mas.

HIP 12843: Despite  large acceleration of 26\,mas/y$^2$, no  RV variability was
    	   found in the GCS from 2 measures.


HIP  14527: The  companion with  $q=0.29$ is  resolved with  NICI but
     unresolved with speckle, being too faint in the optical.

HIP 16370:  The double-lined SB detected  in GCS is
    resolved by  speckle. Its estimated  orbital period 4\,y  explains the
    acceleration.

HIP 16853: Likely triple: the inner pair produces RV variability and acceleration, the outer 
    	   companion at 2\farcs7 is discovered with NICI.
 
HIP 17895: Triple, the RV amplitude of 9.5\,km/s and double lines seen
    in GCS cannot be caused by the visual system YR~23 with estimated
    period  of  $\sim$50\,y  which  could  however  be  responsible  for
    $\Delta \mu$.

HIP 21008: Member of the Hyades (vB~81) and ``probable SB'' according
      to \citet{Griffin88}. This WDS pair PAT~10 was resolved by speckle.

HIP  21053: Hyades. The  0\farcs3 binary  PAT 11  was not  resolved by
     speckle.  The GCS did  not detect  RV variability  despite large
     scatter of 4.8\,km/s in their 3 measures.

HIP 21079: Possibly triple: the new companion at 1\farcs6 should not produce RV variability by 3.8\,km/s.

HIP  21543: Hyades. The  visual companion  CHR~153 at  0\farcs54 shows
    	   only linear  motion \citep{HTM11},  it can be  another
    	   member  of  Hyades  or  a  wide pair  in  projection.   The
    	   acceleration could be produced by the 1.8\,y SB companion.

HIP 21778: Resolved with NICI, uncertain measure. Not resolved with speckle.

HIP 22221: Hyades. The 0\farcs1 pair PAT 16 was not resolved with speckle; its estimated period is 7\,y.

HIP 22387:  The companion is close  to the detection limit,  but it is
    	   considered to be real.  The measurement is uncertain.

HIP 24336: The 1\farcs25 companion discovered with NICI is hardly the one that produced the 
    	   acceleration of 17\,mas/y$^2$.

HIP 24419:  The acceleration  is due to  the 2.2\,y SB  with estimated
    	   axis of 55mas and $q>0.14$.  The companion is too faint and
    	   close to  be resolved with  NICI or speckle. The  system is
    	   triple with a faint common-proper-motion (CPM) companion at 14\farcs6.

HIP 25148: The companion at 66\,mas is at the diffraction limit, the NICI
    	   measurement is not accurate. The pair can be resolved by speckle.

HIP  25180: Large  RV amplitude  (31.7\,km/s) hints  at  short orbital
     period, but  only small  $\Delta \mu =  3$\,mas/y is  detected by
     Hipparcos. Triple?

HIP  28241:  A triple system  consisting  of  the  0\farcs54 inner  pair
resolved with NICI and the physical companion B at 11$''$.

HIP 29860: The 27-y pair  CAT 1 with computed visual \citep{HTM11} and
    	   SB orbits is responsible for $\Delta \mu$.  Triple with the
    	   CPM companion LEP 24 AE at 103$''$.

HIP 37853:  Triple system at  15\,pc. The inner astrometric  pair with
    	   large $\Delta  \mu$ is resolved with speckle.  It is accompanied by 
    	   the white dwarf NLTT 18141 = GJ 288B at 914$''$.

HIP 44874: Possibly  triple: the known companion RST  2610 at 1\farcs7
           with  estimated  period  500\,y  is unlikely  to  cause  the RV
           variability and $\Delta \mu$.

HIP 50870: The 1\farcs1 speckle companion  can explain $\Delta \mu$, but not the acceleration.

HIP  53217: Triple  with inner  SB1  of 6.8d  and outer  SB system  of
    	   estimated 3.3\,y period  which produces large acceleration.
    	   The estimated axis is 40\,mas, unresolved with speckle.

HIP 55714:  The 0\farcs1 pair  CHR 242 has  only 1 observation  in the
    WDS,  unresolved with  speckle. Double  lines were noted  by  the GCS.
    This star remains a mystery.

HIP 64219: Triple: the inner SB1 of 20.4\,d period has tertiary companion TOK 28
    	   at 0\farcs31 with mass ratio of 0.2 which produces the acceleration and $\Delta \mu$.
	   This tertiary is too faint to be resolved in the optical.

HIP 67620: The  10.3-y SB1 orbit by \citet{Abt06} matches the
    	    speckle pair WSI~77 and explains the large $\Delta \mu$

HIP 73241: The 14.9-y SB1 orbit is mentioned by \citet{Raghavan10}.

HIP  92103: The NICI images  are of  poor quality  resulting in  a shallow
     detection limit (at 0\farcs2, $\Delta K <3$ and $\Delta H <2$)

HIP  97676:  NICI  resolved   the  astrometric  binary  with  variable
     	     RV. Triple system with a  CPM companion at 83$''$.

HIP 103260: The  known 3\farcs9 pair I 18  with estimated period $\sim
    1700$\,y cannot explain the acceleration.

HIP 103987: The 1-y SB with large acceleration has been resolved with speckle.
    	    However, we can't exclude that it is a triple system. 

HIP 108095:  NICI images  contain a hint  of faint companion  at $\sim
             280^\circ$, 0\farcs12, not accepted as real.

HIP 109067: The 4.6\,y SB1 with estimated $q>0.23$ and axis 53\,mas is below the NICI detection limit.
    	    This is a sub-dwarf with large PM, below the Main Sequence. 

HIP 109443: Possibly triple. The 1\farcs4 companion found with NICI can't explain the RV variability 
    	    and acceleration.

HIP 114313: SB1 with P=3.1\,y, estimated $q>0.15$. The orbital axis is
     31\,mas, too close for NICI. Large acceleration.


HIP 115505: $\Delta \mu$ is explained by the newly resolved 0\farcs5 pair.
    Triple with  companion at 13\farcs8  in 2MASS.  This  companion is
    measured  with NICI  as  well  at the  same  position, its  colors
    matching a low-mass dwarf.   Considering the low density of background
    stars, the companion is physical, although we cannot confirm it as
    CPM owing to the small PM of the main target.

\end{scriptsize}



\begin{deluxetable}{cc | cc | cc}
\tabletypesize{\scriptsize}
\tablecaption{Detection limits $\Delta m(\rho)$ \label{tab:detlim}}
\tablewidth{0pt}                                                                               
\tablehead{\multicolumn{2}{c|}{Hipparcos} & \multicolumn{2}{c|}{Speckle} &\multicolumn{2}{c}{NICI} \\
$\rho$ & $\Delta V$ & $\rho$ & $\Delta I$ & $\rho$ & $\Delta K$ }
\startdata
0.09 & 0   &  0.04  & 2   &  0.058 & 1 \\
0.14 & 2.2 &  0.15  & 4.0 &  0.144 & 4  \\
0.4  & 4.0 &  0.5   & 5.7 &  0.27  & 7.4 \\
10   & 4.3 &  1.5   & 5.7 &  9.0  & 7.4 \\ 
\enddata
\end{deluxetable}

\begin{deluxetable}{lrrr r rr}
\tabletypesize{\scriptsize}
\tablecaption{Selected distant Hipparcos stars with large accelerations and significant
differences between Tycho-2 and Hipparcos proper motions. \label{tab:acc}}
\tablewidth{0pt}
\tablehead{
HIP & Parallax & HIP PM & Tycho-2 PM & HIP2 PM &  $\dot{\mu}$ & $\dot{\mu}/\sigma(\dot{\mu})$\\
    & mas      &  mas yr$^{-1}$ & mas yr$^{-1}$ & mas yr$^{-1}$ & mas yr$^{-2}$ & }
\startdata
2777 & 1.7 & (46.9, -8.9) & (30.2, -4.8)      & (34.1,-6.2)   & 17.2 & 4.8\\
14133 & 1.4 & (24.2, -17.8) & (3.0, -3.2)     & (1.1,-0.7)    & 25.8 & 4.0\\
55442 & 0.5 & (-25.0, -37.6) & (-34.7, -23.9) & (-31.0,-25.5) & 16.8 & 3.6\\
58036 & -0.1 & (-107.1, -9.3) & (-50.8, 13.0) & (-46.6,12.3)  & 60.6 & 6.2\\
76245 & 1.3 & (-4.9, -2.4) & (-15.6, 1.0)     & (-11.6,-1.6)  & 11.3 & 4.7\\
101344 & 1.2 & (18.5, -63.2) & (11.0, -72.9)  & (12.8,-74.3)  &  12.3 & 4.0\\
101941 & 1.3 & (-14.5, -23.3) & (-18.7, -24.5)& (-15.6,-24.2) &  4.3 & 3.5\\
107466 & 0.2 & (12.3, 21.8) & (12.0, 14.3)    & (16.5,17.7)  & 7.5 & 4.8\\
110978 & 0.9 & (-11.8, -17.2) & (-0.7, -18.4) & (-4.4,-18.8) & 11.1 & 3.6\\
111759 & 1.7 & (-0.9, 0.8) & (-10.8, 3.1)     & (-10.3,3.8)  & 10.2 & 4.3\\
111835 & -1.4 & (-16.0, -24.0) & (-11.0, 4.1) & (-10.2,1.3) & 28.5 & 6.7\\
\enddata
\end{deluxetable}

\end{document}